\documentclass {jaa}
\usepackage[english]{babel}

\def\k{{\bf k}}
\def\kp{k_\parallel}
\def\kpr{{\bf k}_\perp}

\def\HI{{\rm HI}}

\def\u{\vec{U}}

\def\kperp{k_{\perp}}
\def\kpar{k_{\parallel}}

\def\mnras{MNRAS}
\def\jcap{JCAP}

\def\aap{A \& A}
\def\apj{Ap.J}

\def\u{{\bf U}} 
\def\V{\mathcal{V}}
\def\N{{\mathcal N}}
\def\S{{\mathcal S}}

\def\lsim{~\rlap{$<$}{\lower 1.0ex\hbox{$\sim$}}}

\def\gsim{~\rlap{$>$}{\lower 1.0ex\hbox{$\sim$}}}

\usepackage{graphics}
\usepackage{graphicx}
\usepackage{color}
\usepackage{amsmath}
\usepackage{psfrag}
\usepackage{epsfig}

\begin{document}
\title[Fisher matrix predictions for the  $21\, {\rm cm}$ signal with  OWFA ]
{Fisher matrix predictions for detecting the cosmological $21\, {\rm cm}$ signal with  the Ooty Wide Field Array
(OWFA)}
\author[ S. Bharadwaj, A.K. Sarkar and S. S. Ali] {Somnath Bharadwaj$^{1,2}$\thanks{Email:somnath@phy.iitkgp.ernet.in}, Anjan Kumar Sarkar$^{2}$\thanks{Email:anjan@cts.iitkgp.ernet.in} and Sk. Saiyad Ali$^{3}$\thanks{Email:saiyad@phys.jdvu.ac.in}\\ 
    $^{1}$ Department of Physics,  IIT Kharagpur, 721302, India \\
    $^{2}$ Centre for Theoretical Studies, IIT Kharagpur, 721302 , India \\
    $^{3}$ Department of Physics, Jadavpur University, Kolkata 700032, India}
\date {}
\maketitle

\begin{abstract}
We have used the Fisher matrix formalism  to quantify the prospects of detecting  the  $z = 3.35$ 
redshifted 21-cm  HI power spectrum  with the upcoming radio-imterferometric array OWFA. 
OWFA's  frequency and baseline  coverage spans  comoving Fourier modes in the range 
$1.8  \times 10^{-2} \le  k \le 2.7 \, {\rm Mpc}^{-1}$. 
The OWFA HI signal, however,  is predominantly from the range $k \le 0.2  \, {\rm Mpc}^{-1}$. The larger modes, though 
abundant, do not contribute much to the HI signal. 
In this work we  have focused on combining 
the entire signal to achieve a detection. We find that a $5-\sigma$ detection of $A_{HI}$ is possible 
with $\sim 150 \, {\rm hr}$ of observations, here $A_{HI}^2$ is the amplitude of the HI power spectrum. 
We have also carried out a joint analysis for $A_{HI}$ and $\beta$ the redshift space distortion parameter. 
Our study shows that OWFA is very sensitive to the amplitude of the HI power spectrum.
However, the anisotropic distribution of the $\k$ modes does not make it very suitable 
for measuring $\beta$.

\end{abstract}


\section{Introduction}    
Work is currently in progress to upgrade the cylindrical Ooty Radio
Telescope (ORT\footnote{http://rac.ncra.tifr.res.in/})) so that it
functions as a linear interferometric array the Ooty Wide Field Array
(OWFA; Prasad \& Subrahmanya 2011a,b; Ram Marthi \& Chengalur 2014).
This telescope operates at a nominal frequency of $\nu_o= 326.5 \,{\rm
MHz}$ which corresponds to the neutral hydrogen (HI) $1,420 \, {\rm MHz}$ 
radiation from a  redshift $z=3.35$. Observations of the fluctuations 
in the contribution from the HI   to the diffuse background radiation are
a very interesting  probe of the large-scale structures in the high-$z$
universe  (Bharadwaj, Nath \& Sethi 2001,Bharadwaj, \& Sethi 2001). In addition
to the power spectrum (Bharadwaj \& Pandey 2003, Bharadwaj \& Srikant 2004)  
this is also a sensitive probe of the bispectrum (Ali, Bharadwaj and Pandey 2006, Guha Sarkar  \& Hazra  2013). There has been a continued, 
growing interest towards  the detection of the 21 cm signal
from the lower redshifts $(0 < z < 4)$ to probe the post-reionization
era (Chang et al. 2008;  Visbal et al. 2009;
Bharadwaj et al. 2009; Wyithe \& Loeb 2009; Bagla, Khandai \& Datta
2010; Seo et al. 2010; Mao 2012; Ansari et al. 2012; Bull et
al. 2014; Villaescusa-Navarro et al. 2014). Recently, Ali \& Bharadwaj (2014) (henceforth, Paper I)
have studied the prospects for  detecting the HI signal from 
redshift $z=3.35$ using  OWFA. The OWFA provides an unique opportunity 
to study the large scale structures  at $z=3.35$.
 
A number of similar upcoming packed radio interferometer (CHIME\footnote{http://chime.phas.ubc.ca/}, Bandura et al. 2014;
BAOBAB\footnote{http://bao.berkeley.edu/} and  the KZN array\footnote{A
  compacted array of 1225 dishes with diameter 5m  each, 
based on BAOBAB and sited in South Africa}) have been proposed
to probe the expansion history of the low-redshift universe ($z \le
2.55$) with an unprecedented precision using BAO measurements from the
large-scale HI fluctuations. Even more innovative designs are being
planned for the future low frequency telescope
SKA\footnote{http://www.skatelescope.org/}. This promises to yield
highly significant measurements of the HI power spectrum over a large
 redshift range spanning nearly the entire post-reionization era $(z < 6)$. 
However, the detection of the faint $21\,\rm cm$ HI signal 
($ \sim 1\, {\rm mK}$) is extremely challenging due to the presence of 
different astrophysical foregrounds. The foregrounds are four to five
 orders of magnitude brighter than the post-reionization HI signal 
(Ghosh et al. 2011a,2011b).

In this paper, we have considered  the visibility correlation 
(Bharadwaj, \& Sethi 2001, Bharadwaj, \& Ali 2005) which 
essentially is the data covariance matrix that is necessary to 
calculate  the Fisher matrix. We have employed the Fisher matrix technique
to predict the expected signal-to-noise ratios (SNR) for detecting the
HI signal. In our analysis we have assumed that the HI traces the 
total  matter with a linear bias, and the  matter power spectrum is 
precisely as predicted by the standard LCDM model with the parameter 
values mentioned later. The HI power spectrum is then completely 
specified by two parameters  $A_{HI}$, which sets the overall  amplitude
of the power spectrum, and $\beta$ the redshift distortion parameter. 
The parameter $A_{HI}$ here is  the product of the mean
neutral hydrogen fraction ($\bar{x}_{\HI}$) and the linear bias
parameter ($ b_{HI}$). For a detection, we focus on measuring 
the amplitude  $A_{HI}$, marginalizing over $\beta$. We also 
consider the joint estimation of  $A_{HI}$ and $\beta$.
Our entire analysis is based on the assumption that the 
visibility data contains only the signal and the noise, and 
the foregrounds and radio-frequency interference have been completely
removed from the data.

The BAO feature is within the baseline range covered by OWFA (Paper I). 
However, the frequency coverage ($\sim 30 \, {\rm MHz}$) is rather small. 
Further, for the present analysis we have only considered 
observations in a single field of view.  All of these result in 
having very few Fourier modes across the $k$ range relevant for the 
BAO, and we do not consider this here.

The rest of the paper is organized as follows.  Section 2 briefly 
discusses some relevant system parameters for OWFA. 
In Section 3, we present the theoretical model for calculating the
signal and noise covariance,  and predict their respective contributions.
 Here we also estimate the range of k-modes which are probed by the
OWFA. In Section 4 we use  the Fisher matrix analysis to make 
predictions for  the SNR  as a function of the observing time. 
Finally, we present summary and conclusions in
Section 5.

In this paper, we have used the (Planck $+$ WMAP) best-fit
$\Lambda$CDM cosmology with cosmological parameters (Ade et al. 2013):
$\Omega_{m}=0.318, \Omega_bh^2=0.022,\Omega_{\Lambda}=0.682,
n_s=0.961, \sigma_8=0.834, h=0.67$. We have used the matter transfer
function from the fitting formula of Eisenstein \& Hu (1998)
incorporating the effect of baryonic features.

\section{Telescope parameters}

\begin{table}
\begin{center}
\caption{System parameters for Phases I, II, III and IV of the OWFA .}
\vspace{.2in}
\label{tab:array}
\begin{tabular}[scale=.3]{|l|c|c|c|c|}
\hline \hline Parameter & Phase I &Phase II& Phase III & Phase IV\\ 
\hline No. of antennas & 40 & 264 & 528 &1056 \\ 
($N_A$)&  & & &\\
\hline No. of dipoles $N_d$ & 24 & 4& 2 & 1 \\
\hline Aperture area  & $30
\,{\rm m} \times 11.5 \,{\rm m}$ & $ 30 \,{\rm m} \times 1.92 \,{\rm
  m} $ & $ 30 \,{\rm m} \times 0.96 \,{\rm
  m} $ & $ 30 \,{\rm m} \times 0.48 \,{\rm m} $\\ 
($b \times d$)&  & & &\\

\hline Field of View & $ 1.75^{\circ} \times
4.6^{\circ}$ & $ 1.75^{\circ} \times 27.4^{\circ}$ &  $ 1.75^{\circ} \times 54.8^{\circ}$& $ 1.75^{\circ} \times 109.6^{\circ}$ \\ 
(FoV)&  & & &\\

\hline Smallest
baseline  & $11.5 \,{\rm m} $ & $1.9 \,{\rm m} $&$0.96 \,{\rm m} $ &$0.48 \,{\rm m} $ \\ 
($d_{min}$) &  & & &\\

\hline
Largest baseline  & $ 448.5 \,{\rm m}$ & $505.0 \,{\rm
  m}$&  $506.0 \,{\rm m}$&  $506.5 \,{\rm m}$ \\
 ($d_{max}$)&  & & &\\

\hline Total band- & $18 \,{\rm MHz}$ & $30 \,{\rm MHz}$&$60 \,{\rm MHz}$ &$120 \,{\rm MHz}$ \\ width (B) &  & & &\\

\hline Single Visibility  & $1.12$ Jy & $6.69 $ Jy  & $13.38 $ Jy  & $26.76 $ Jy \\  rms. noise ($\sigma$) &  & & &\\ \hline
\end{tabular}
\end{center}
\end{table}

The ORT is   a 530 m long and 30 m wide parabolic cylindrical reflector placed in
 the north-south
direction on a hill with the same slope as the latitude $(11^{\circ})$ of the station
(Swarup et al. 1971; Sarma et al. 1975). It thus becomes possible to observe
the same part of the sky by rotating the parabolic cylinder along its long
axis.The telescope has  1056 half-wavelength $ (0.5 \lambda_0 \approx  0.5 {\rm m})$ dipoles 
 placed nearly end to end
 along the focal line of the cylinder. Work is underway to implement electronics that 
combines  the digitized  signal from every  $N_d$  successive dipoles so that we have a linear
array of $N_A$ antennas located along the length of the cylinder.  The OWFA will, at present, 
 have the ability to operate  in two different modes one with $N_d=24$ and 
another with $N_d=4$, referred to as Pase I and Phase II respectively. For our theoretical 
analysis we have also considered two hypothetical (possibly future) upgrades Phases III 
and IV with $N_d=2$ and $N_d=1$ respectively.  Table \ref{tab:array} summarizes various 
parameters for different phases of the array. 
The individual antennas get more compact,  and the field of view increases  from Phase I to IV.   
The  number of antennas and  the  frequency bandwidth also increases  from Phase I to IV.

For any phase, each antenna has a rectangular aperture of dimensions $b\times d$,  and is
distributed at an interval  ${\bf d}=d \, {\bf \hat{i}}$ along the length of the cylinder. 
The value of $b(=30 {\rm m})$,  which corresponds to the width of the parabolic reflector, 
remains fixed for all the phases. The value of $d$ varies for the different phases (Table \ref{tab:array} ). 
The baseline $\u$ quantifies  the antenna pair separation
projected perpendicular to the line of sight measured in
the units of the observing wavelength $\lambda$. Assuming observations 
vertically overhead, we have the baselines 
\begin{equation}
\u_a = a \frac{{\bf d}}{\lambda}  \hspace{2.0cm} (1 \le a \le N_A-1) \,.
\end{equation}

In reality
$\u_1,\u_2,...$ vary across the observing bandwidth as frequency
changes.  However, for the present purpose of the paper we keep $\u_a$
fixed at the value corresponding to the nominal frequency.

A schematic view of the OWFA array layout
 is presented in  Paper I. The OWFA has a 
significant number of redundant baselines. For any baseline $\u_a$  we
have $M_a = (N_A - a)$ times sampling redundancy of the baseline.

\section{OWFA visibility covariance  and the Fisher matrix}
\label{sec:vc}
The OWFA  measures  visibilities $\V(\u_a,\nu_m)$ 
at a finite number of baselines $\u_a$ and frequency channels $\nu_m$  
with frequency channel width  $\Delta \nu_c$ 
spanning a frequency bandwidth $B$.
The measured   visibilities  can be expressed as a 
combination of the HI signal and the noise
\begin{equation}
\V(\u_a,\nu_m)=\S(\u_a,\nu_m)+\N(\u_a,\nu_m)
\label{eq:b1}
\end{equation}
assuming that  the foregrounds have been  removed. 
The correlation expected between the  HI  signal 
at two different baselines and frequencies can be 
calculated (Paper I and references therein) using 
\begin{eqnarray}
\langle \S(\u_a,\nu_n)  \S^{*}(\u_b,\nu_m) \rangle &=& 
\left(\frac{2 k_B}{\lambda^2}\right)^2
\int d^2 U^{'} \tilde{A}(\u_a-\u^{'}) 
\tilde{A}^{*}(\u_b-\u^{'}) \nonumber \\
&\times&  \frac{1}{2 \pi r_{\nu}^2} 
\int d \kp \cos(\kp r_{\nu}^{'} \Delta \nu )
 P_{\rm HI}(\frac{2 \pi \u^{'}}{r_{\nu}},\kp)
\label{eq:a3}
\end{eqnarray}  
where $P_{\rm HI}(\kpr,\kp)$ is the power spectrum of the 21-cm 
brightness temperature fluctuation in redshift space, 
$\left(\frac{2 k_B}{\lambda^2}\right)$ is the conversion from 
brightness temperature to specific intensity,  $r_{\nu}$ is the
comoving distance from the observer to the region where the HI
radiation originated, $r_{\nu}^{'}=dr/d \nu $ is the radial
conversion factor from frequency interval  to comoving separation ($r_{\nu}=6.85$ Gpc and $r_{\nu}^{'}=11.5$ Mpc MHZ$^{-1}$ for OWFA),
$\Delta \nu=\nu_m-\nu_n$ and   $\tilde{A}(\u)$ is the Fourier 
transform of the OWFA primary beam pattern. 

The real and imaginary parts of the noise $\N(\u_a,\nu_n)$
both have equal variance
$\sigma^2$ with  
\begin{equation}
\sigma
=\frac{\sqrt{2}k_BT_{sys}}{\eta A\sqrt{\Delta \nu_c t}}
\label{eq:a4}
\end{equation}
where $T_{sys}$ is the system Temperature, $\eta$ and $A=b \times d$ are 
respectively the efficiency and the geometrical area of the antenna aperture
and $t$ is the observing time. 
We have used the values $ T_{sys}=150 \, {\rm K}$, $\eta=0.6$ and  $\Delta \nu_c=0.1 \, {\rm MHz}$
which are the same as in Paper I. 
 
The noise in the visibilities measured at different baselines and 
frequency channels are uncorrelated. 
We then have 
\begin{equation}
\langle \N(\u_a,\nu_n) \N^{*}(\u_b,\nu_m) \rangle =\delta_{a,b}
\delta_{n,m} 2 \sigma^2 \,.
\label{eq:a5a}
\end{equation}

Earlier studies (Paper I)
have shown that for a fixed baseline ($U_a=U_b$) the HI signal 
(eq.~\ref{eq:a3})
is correlated out to frequency separations $\mid \nu_n -\nu_m \mid \sim 0.5 \, {\rm MHz}$
which spans  several  frequency channels. This implies that the 
data covariance matrix $\langle \V(\u_a,\nu_n) \V^{*}(\u_a,\nu_m) \rangle$ 
has considerable off-diagonal terms, a 
feature that is not very convenient for the Fisher matrix analysis.

For the Fisher Matrix analysis it is convenient to use  
the delay channels $\tau_m$ (Morales 2005) instead of the frequency 
channels $\nu_c$. We define  
\begin{equation}
v(\u_a,\tau_m)=\Delta \nu_c \sum_n  e^{2 \pi i \tau_m  (\nu_n-\nu_0)} \V(\u_a,\nu_n)
\label{eq:b4}
\end{equation}
where $N_c$ is the number of frequency channels, 
$B= N_c  \Delta \nu_c$ and 
$$\tau_m=\frac{m}{B}  \hspace{1.3cm} \frac{-N_c}{2} < m \le \frac{N_c}{2}\, . $$
The covariance matrix $\langle  v(\u_a,\tau_m) v^*(\u_b,\tau_n) \rangle $ is zero if 
$n \neq m$, and we need only consider the diagonal terms $n=m$. Defining 
$C_{ab}(m)=\langle  v(\u_a,\tau_m) v^*(\u_b,\tau_m) \rangle $ we have 

\begin{eqnarray}
C_{ab}(m)  &=& \frac{B}{
  r_{\nu}^2 r_{\nu}^{'}} \left(\frac{2 k_B}{\lambda^2}\right)^2 
\int d^2 U^{'}
\tilde{A}(\u_a-\u^{'})    \tilde{A}^{*}(\u_b-\u^{'}) 
P_{\rm HI}(\frac{2 \pi \u^{'}}{r_{\nu}},\frac{2 \pi \tau_m}{r_{\nu}^{'}})
\nonumber \\
 &+& \delta_{a,b} \,  2 \,  \Delta \nu_c  \, B  \,   \frac{\sigma^2}{(N_A - a)} \,.
\label{eq:a5}
\end{eqnarray}
The factor $(N_A - a)^{-1}$ in the noise contribution accounts 
for the redundancy in the baseline distribution. 
The functions $\tilde{A}(\u_a-\u^{'})$ and  $\tilde{A}^{*}(\u_b-\u^{'})$
have an overlap only if $a=b$ or $a=b \pm 1$ (Paper I). The visibilities
at two baselines$\u_a$ and $\u_b$  are uncorrelated $(C_{ab}(m)=0)$
if $\mid a -b \mid > 1$ {\it ie.} the visibility at a particular baseline
$\u_a$  is only correlated with the other visibility measurements at the 
same baseline or the adjacent baselines $\u_{a \pm 1}$.   Thus, for a 
fixed $m$, $C_{ab}(m)$ is a symmetric, tridiagonal matrix.
Further, the noise  only contributes to the diagonal terms, and it does 
not figure in the off-diagonal terms.

We use the data covariance $C_{ab}(m)$to 
calculate the Fisher Matrix using 
\begin{equation}
F_{\alpha \beta}=\frac{1}{2} \sum_m C^{-1}_{ab}(m)
[C_{bc}(m)]_{,\alpha}  C^{-1}_{cd}(m) [C_{da}(m)]_{,\beta}  
\label{eq:a6}
\end{equation}
where the indices $a,b,c,d$ are to be summed over all baselines, 
and $\alpha,\beta$ refer to the various parameters which are to be
estimated from the OWFA data.

It is possible to get further  insight into the cosmological information 
contained in  the data covariance $C_{ab}(m)$ by considering large 
baselines $U_a \gg d/\lambda$ where it is reasonable to assume that the 
function  $\tilde{A}(\u_a-\u^{'})    \tilde{A}^{*}(\u_b-\u^{'}) $
in eq.~(\ref{eq:a5}) falls  sharply in comparison to the slowly changing 
HI power spectrum  as $\u^{'}$ is varied. The integral in equation 
eq.~(\ref{eq:a5}) can then be approximated as 
\begin{equation}
\approx P_{\rm HI}(\k) 
\int d^2 U^{'}
\tilde{A}(\u_a-\u^{'})    \tilde{A}^{*}(\u_b-\u^{'}) 
\label{eq:a7}
\end{equation}
where 
\begin{equation}
\k \equiv (\kpr,\kp)
\equiv (\pi[\u_a+\u_b]/r_{\nu},2 \pi \tau_m/r_{\nu}^{'})\,.
\label{eq:a8}
\end{equation}
The integral in eq.~(\ref{eq:a7}) can be evaluated analytically, and we 
have the approximate formula  
\begin{eqnarray}
C_{ab}(m)  = B \left[ 
\frac{(2 k_B)^2 (4 \delta_{a,b}+\delta_{a,b \pm 1})}
{ 9 \lambda^2 b d r_{\nu}^2 r_{\nu}^{'}} 
P_{\rm HI}(\k)+  \frac{\delta_{a,b} \,  2 \,  \Delta \nu_c  
\sigma^2}{(N_A - a)} \,\right] \,.
\label{eq:a9}
\end{eqnarray}

\begin{figure}[ht]
\begin{center}
\psfrag{Signal Covariance}[c][c][1.0][0]{$S_{ab}(m)$\ Jy$^{2}$MHZ$^{2}$}
\psfrag{p1}[c][c][0.8][0]{$(a=b)$}
\psfrag{p2}[c][c][0.8][0]{$(a=b\pm1)$}
\psfrag{PHASE II}[c][c][1.1][0]{PHASE II}
\includegraphics[scale =.5]{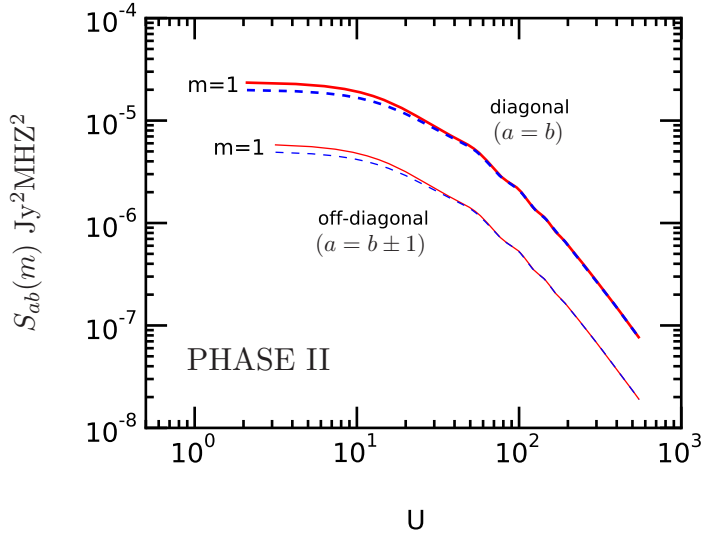}
\caption{This shows the signal contribution to the covariance matrix $C_{ab}(m)$  
for $m=1$  calculated using eq.  (\ref{eq:a5}) (solid curves) and the approximate formula 
eq. (\ref{eq:a9}) (dashed curves).}
\label{fig:convo2}
\end{center}
\end{figure}

Figure \ref{fig:convo2} shows a comparison of  the signal contribution to the 
covariance matrix calculated using eq.  (\ref{eq:a5}) and the approximate formula 
eq. (\ref{eq:a9}).  We find that the results are in reasonably good agreement over  the 
entire $U$ range for $m=1$. The agreement is better at large baselines $U \ge 30$ where the 
two curves are nearly undistinguishable. The results are  indistinguishable  for the entire 
$U$ range for $m>1$  which has not been shown here. Although we have used the 
approximate equation (eq. \ref{eq:a9}) to interpret $C_{ab}(m)$  in the subsequent discussion, we have 
used  eq. (\ref{eq:a5}) to compute $C_{ab}(m)$ throughout the entire analysis. 

Returning to eq. (\ref{eq:a9}),
first, the signal contribution to the diagonal terms is found to  be 
$4$ times larger than the off-diagonal terms. Next, 
we see that each non-zero element of the covariance matrix $C_{ab}(m)$  
corresponds to the HI power spectrum at a particular  comoving Fourier mode $\k$ 
given by eq.~(\ref{eq:a8}). Each  delay channel $\tau_m$  corresponds to a 
$k_{\parallel m}=2 \pi \tau_m/r_{\nu}^{'}$ which  spans the values 
\begin{equation}
k_{\parallel m}=m \left( \frac{2 \pi }{B r_{\nu}^{'}} \right) 
 \hspace{1cm} \frac{-N_c}{2} < m \le \frac{N_c}{2}\, .
\end{equation}
For a fixed $\tau_m$,  the diagonal terms of $C_{ab}(m)$ 
 with $\u_a=\u_b$ correspond to  $k_{\perp a}= 2 \pi U_a/r_{\nu}$
which spans the values 
\begin{equation}
k_{\perp a}= a \left( \frac{2 \pi d}{\lambda r_{\nu}} \right) \hspace{1cm} 
1 \le a \le N_A-1 \,,
\end{equation}
and the  off-diagonal terms of $C_{ab}(m)$ with $\u_b=\u_{a+1}$ correspond to
$k_{\perp a}= \pi [U_a+U_b]/r_{\nu}$ 
which spans the values 
\begin{equation}
k_{\perp a}= (a \pm 0.5) \left( \frac{2 \pi d}{\lambda r_{\nu}} \right)  \hspace{1cm} 
1 \le a \le N_A-2 \,,
\end{equation}
We see that the $k_{\perp}$ value probed by any  off-diagonal term is  located 
mid-way between the  $k_{\perp}$  values probed by the two nearest diagonal terms. 
Considering  both the diagonal and the off-diagonal terms, we find that 
the different  $k_{\perp}$ values that will be probed by OWFA are located at an 
interval of $\Delta k_{\perp}= \pi d/(\lambda r_{\nu})$.

\begin{table}
\begin{center}
\caption{The  $k_{\perp}$ and $k_{\parallel}$ range that will be probed 
by the different Phases of  OWFA.}
\vspace{.2in}
\label{tab:kperp}
\begin{tabular}[scale=.3]{|l|c|c|c|c|}
\hline \hline ${\rm Mpc^{-1}}$ & Phase I &Phase II& Phase III & Phase IV\\ 
\hline ${k_{\perp}[min]}$\,  & $1.1\times 10^{-2}$ & $ 1.9\times 10^{-3} $ & $ 9.5\times 10^{-4} $ & $4.8\times 10^{-4}  $ \\ 
\hline  ${k_{\perp}[max]}$  & $4.8\times 10^{-1}$ & $ 5.0\times 10^{-1} $ & $ 5.1\times 10^{-1} $ & $5.1\times 10^{-1} $ \\
\hline ${k_{\parallel}[min]}$  & $3.0\times 10^{-2}$ & $ 1.8\times 10^{-2} $ & $ 9.1\times 10^{-3} $ & $4.6\times 10^{-3}  $ \\ 
\hline  ${k_{\parallel}[max]} $  & $2.73 $ & $2.73 $ & $ 2.73$ & $2.73 $ 
 \\ \hline
\end{tabular}
\end{center}
\end{table}
In addition to the HI signal and the noise considered in this paper, the OWFA visibilities
will also contain a foreground contribution. For the purpose of this work we make the 
simplifying assumption that the  foregrounds are constant across  different frequency channels,
and hence they only contribute to the $\kp=0$ mode. In reality the foreground contamination 
will possibly extend to other modes also. However, in this work we make  the most optimistic
assumption that the  foregrounds will be restricted to the   $\kp=0$ mode and we have excluded 
this in the subsequent analysis. Table~\ref{tab:kperp} shows  the $k_{\perp},k_{\parallel}$ 
range that will be probed by the different Phases of OFWA. We see that for all the phases 
(except Phase I)  the minimum value of $\kp$ is approximately $10$ times larger than the corresponding 
$\kperp[min]$.  The sampling along  $\kp$, which is decided by $1/B$,
has a spacing $\Delta \kp=\kp[min]$ which 
also  is $\sim 5$ times larger than $\Delta \kperp=\kperp[min]/2$ which is decided by the antenna spacing 
$d$. The maximum  $\kp$ values also are approximately $4$ larger than the  corresponding $\kperp[max]$. 
It is thus clear that the sampling in $\kperp$ is quite different from the $\kpar$ sampling, and the  
$\kpar$  
values are several times  larger than the $\kperp$ values. 
This disparity  in the $\kp$ and $\kperp$ coverage and sampling   poses  a problem for using OWFA to 
quantify redshift space distortion. We shall return to this in Section \ref{sec:sum}
 where we discuss the results of our analysis.

\begin{figure}[ht]
\begin{center}
\vskip.2cm 
\psfrag{k1}[c][c][1.0][0]{$P_{HI}(k)$\ Mpc$^{3}$ mK$^{2}$}
\psfrag{k2}[c][c][1.0][0]{$k$\ Mpc$^{-1}$}
\psfrag{PHASE II}[c][c][1.1][0]{PHASE II}
\psfrag{z=3.35}[c][c][1.0][0]{z=3.35}
\centerline{\includegraphics[scale =.5]{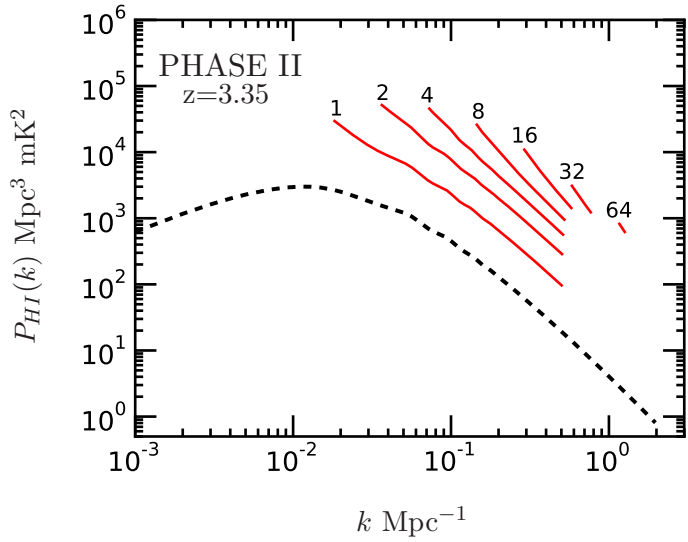}}
\caption{The $k$ range that will be probed by $C_{ab}(m)$ for different
values of $m$. The curves for different $m$ have been arbitrarily
displaced vertically to make them distinguishable.  For reference, 
we have also shown the expected 21-cm  brightness temperature
fluctuation $P_{HI}(k)$  (dashed curve)
where  $P_{HI}(k) \equiv  P_{HI}(k,\mu=0)$ is the  $z=3.35$
HI 21-cm brightness temperature  power spectrum (eq. \ref{eq:pk}).}  
\label{fig:pkHI1}
\end{center}
\end{figure}

\begin{figure}[t]
\begin{center}
\psfrag{Number of k-modes}[c][c][1.0][0]{$\Delta N_k$}
\psfrag{k}[c][c][1.0][0]{$k$\ Mpc$^{-1}$}
\psfrag{PHASE II}[c][c][1.1][0]{PHASE II}
\psfrag{p1}[c][c][0.9][0]{$\Delta k = 0.03$ Mpc$^{-1}$}
\vskip.2cm \centerline{\includegraphics[scale =.5]{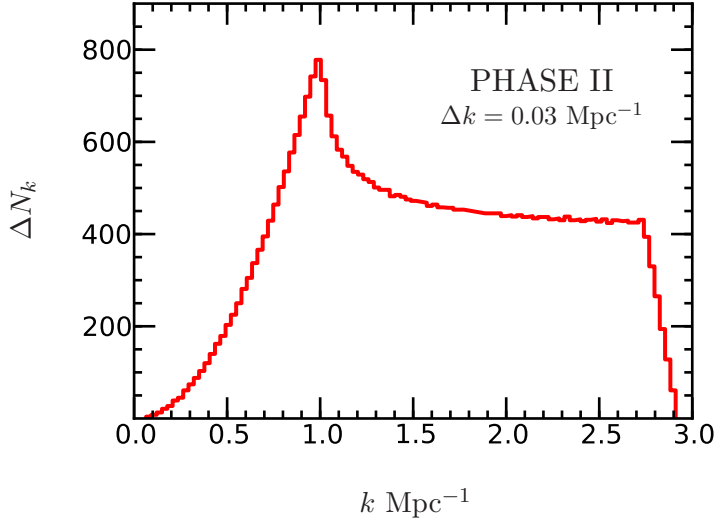}} 
\caption{The histogram shows the number of  $k$ modes, $\Delta N_k$ within bin width $\Delta k$. }
\label{fig:pkHI2}
\end{center}
\end{figure}

Figure \ref{fig:pkHI1} shows the $k=\mid \k \mid =\sqrt{\kp^2 + \kperp^2}$ 
range that will be probed by $C_{ab}(m)$ for different  values of $m$. 
We see that the range $k \sim \kpar[min]$ to $k \sim \kperp[max]$ is probed
for  $m=1$. The $k$ range shifts to larger $k$ values as $m$ is
increased,
and the entire $k$ range lies beyond $1 \, {\rm Mpc}^{-1}$ for $m \ge 64$. 
Figure \ref{fig:pkHI2} shows a histogram of all the different $k$
modes that will be probed by OWFA Phase II.  
We expect the number of modes $\Delta N_k$ in  bins of constant width
$\Delta k$ to scale as $\Delta N_k \sim k^2 \, \Delta k$ if the $\k$ modes
are uniformly distributed  
in three dimensions (3D).  The modes $\k$  have a  2D distribution 
for OWFA,  and we expect   $\Delta N_k \sim k \, \Delta k$ if the
  modes are uniformly distributed. 
However,  we have seen that the distribution is not uniform  
($\Delta \kpar$ and $\Delta \kperp$ have
different values) and the histogram does not show the expected
 linear behaviour. The increase in  $\Delta N_k$
is faster  than linear, it peaks at $k \sim 1 \, {\rm Mpc}^{-1}$ and 
is nearly constant at $\sim 60 \, \%$ 
of the peak value for larger modes out to 
$k \le \kpar[max] \sim \, 3 \, {\rm Mpc}^{-1}$. 
It is  clear that the a very large fraction of the Fourier modes $k$ 
that will 
be probed by OWFA  are in the range $ 1 -3 \, {\rm Mpc}^{-1}$. 
We see that the  Fourier modes all lie in this range for   $m \ge 64$
(Figure \ref{fig:pkHI1}). 
The range $ k < 1 \, {\rm Mpc}^{-1}$ will be sampled by 
a relatively small fraction of the modes,  
and the range $ k < 0.1 \, {\rm Mpc}^{-1}$ will only be sampled 
for $m \le 5$.

\begin{figure}[t]
\begin{center}
\psfrag{Covariance}[c][c][1.0][0]{$C_{ab}(m)$\ Jy$^{2}$ MHZ$^{2}$}
\psfrag{PHASE II}[c][c][1.1][0]{PHASE II}
\psfrag{10hr}[c][c][1.0][0]{10 hr}
\psfrag{100hr}[c][c][1.0][0]{100 hr}
\psfrag{1000hr}[c][c][1.0][0]{1000 hr}
\includegraphics[scale =.5]{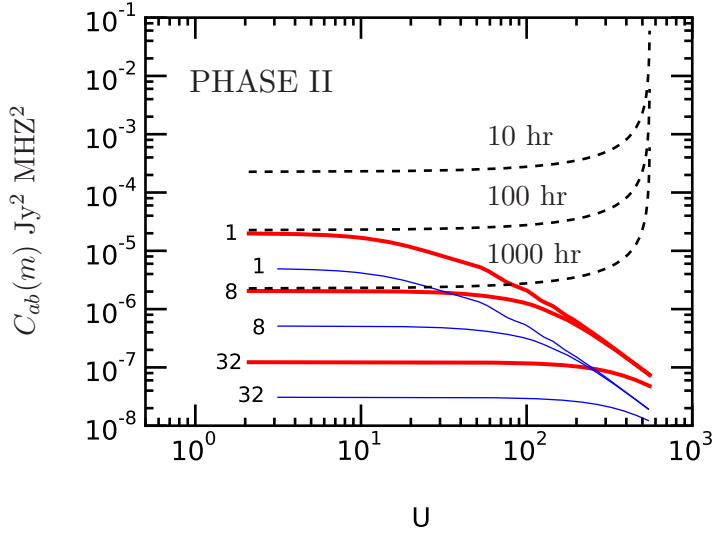}
\caption{This shows the diagonal (thick red curve) and  the off-diagonal (thin blue curve) elements 
of the signal contribution to the covariance  matrix $S_{ab}(m)$  $m= 1$, $8$ and $32$. 
The  system noise contribution  (thick dashed black curves) to $C_{ab}(m)$ is shown for the 
different observing times indicated in the figure.} 
\label{fig:convo1}
\end{center}
\end{figure}

 Figure  \ref{fig:convo1} shows the diagonal  and  the off-diagonal
 elements  of the signal contribution to the covariance matrix $C_{ab}(m)$ 
(eq. \ref{eq:a5}).  The noise contribution is also shown for reference. 
The noise contribution is independent of $m$, and it increases at the larger 
baselines which have a lesser  redundancy $N_{A}-a$. The power spectrum $P_{HI}(k)$ 
is a decreasing function of $k$ for $k \ge 0.1 \, {\rm Mpc}^{-1}$, and most of the 
modes that will be probed by OWFA lie in this range. For a fixed $m$, the signal 
contribution is nearly flat for $U < r_{\nu} m/(B r^{'}_{\nu})$   and then decreases  
if $U$ is increased further.  For $m=1$,  the signal at small baselines $U \le 10$
is comparable to the noise for $T=100 \, {\rm hr}$. The signal is smaller than the noise
 at larger baselines.  The overall amplitude of the signal contribution 
decreases for larger values of $m$,
  The signal covariance falls by a factor of 
$\sim 10$ from $m=1$ to  $m=8$, and it is comparable to the noise for  $T=1,000 \, {\rm hr}$. 
 The signal falls by another factor of $\sim 20$ from $m=8$ if we consider $m=32$. 
 We see that the HI signal is relatively more dominant at the small delay 
channels and  the small baselines.  The HI signal is considerably weaker
at the larger $m$ and $U$, the noise also is  considerably higher at the 
larger baselines.

\section{Results}
\label{fma}

We have assumed that the HI gas, which  is  believed to be 
associated
with galaxies, traces the  underlying  matter distribution with a
constant scale independent  large-scale  linear HI bias $b_{HI}$.
Incorporating redshift space distortion, we have the HI power
spectrum 
\begin{equation}
 P_{\HI}(\k)= A_{HI}^2 \, {\bar{T}^2} \, \left[ 1+ \beta\, {\mu^2} \right]^2
\,P(k) \,.
\label{eq:pk}
\end{equation}
where $P(k)$ is the matter power spectrum,  $\mu= \kp/k$, and 
\begin{equation}
 \bar{T}(z) = 4.66 \, {\rm mK}
\, (1+z)^2\, \left(\frac{\Omega_b h^2}{0.022} \right) \,
\left(\frac{0.67}{h} \right) \, \left( \frac{H_0}{H(z)} \right) \,.
\end{equation}

The parameter   $A_{HI}$ in eq. (\ref{eq:pk})  sets the overall  amplitude
of the HI power spectrum, and $A_{HI}= \bar{x}_{\HI} \, b_{HI}$ where $\bar{x}_{\HI}$
is  the mean neutral hydrogen fraction. The parameter $\beta=f(\Omega)/  b_{HI}$
is the linear redshift distortion parameter. Note that the various terms in 
eq.  (\ref{eq:pk}) are all at the redshift where the HI radiation originated, 
which is $z=3.35$ for the OWFA. 

We have used the value $\bar{x}_{\HI} =0.02$ which corresponds to
$\Omega_{gas}=10^{-3}$ from DLA observations (Prochaska \& Wolfe 2009;
Noterdaeme et al. 2012; Zafar et al. 2013)  in the redshift range of 
our interest. N-body
simulations (Bagla, Khandai \& Datta 2010; Guha Sarkar et al. 2012)
indicate that it is reasonably well justified to assume a constant HI
bias $b_{HI}=2$ at wave numbers $k \le 1 \, {\rm Mpc}^{-1}$, and we have
used this value for our entire analysis. This  is also 
consistant with the Semi-empirical simulations  of 
Mar{\'{\i}}n et al. (2010). Using these values and the cosmological 
parameters values assumed earlier, we have  $A_{HI}=4.0\times 10^{-2}$ and
$\beta=4.93 \times10^{-1}$ which serve as  the fiducial values for our analysis. 

We have assumed that $\bar{T}$ and the $\Lambda$CDM matter power spectrum  $P(k)$ 
are precisely known, and we have used the Fisher matrix analysis to determine the 
accuracy with  which it  will be possible to measure  the parameters $A_{HI}$ and $\beta$
using OWFA observations. The  Fisher matrix analysis (eq. \ref{eq:a6})
was carried out with the two parameters $q_1=\ln(A_{HI})$ and $q_2=\ln(\beta)$. 

\begin{figure}
\begin{center}
\psfrag{SNR}[c][c][1.0][0]{SNR}
\psfrag{T (hours)}[c][c][1.0][0]{t hr}
\vskip.2cm 
\centerline{{\includegraphics[scale =.33]{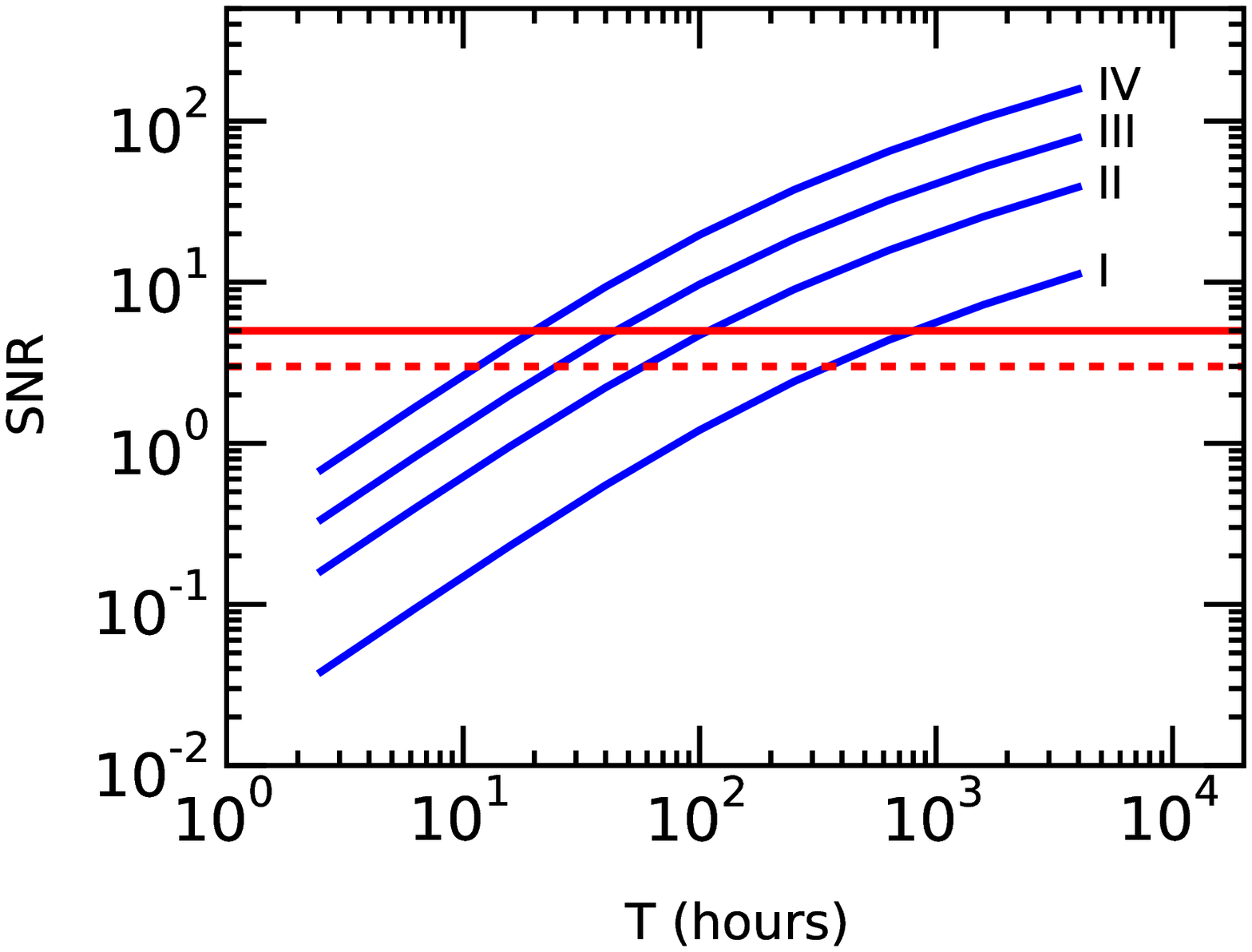}}
  \hskip0.01cm { \includegraphics[scale =.33]{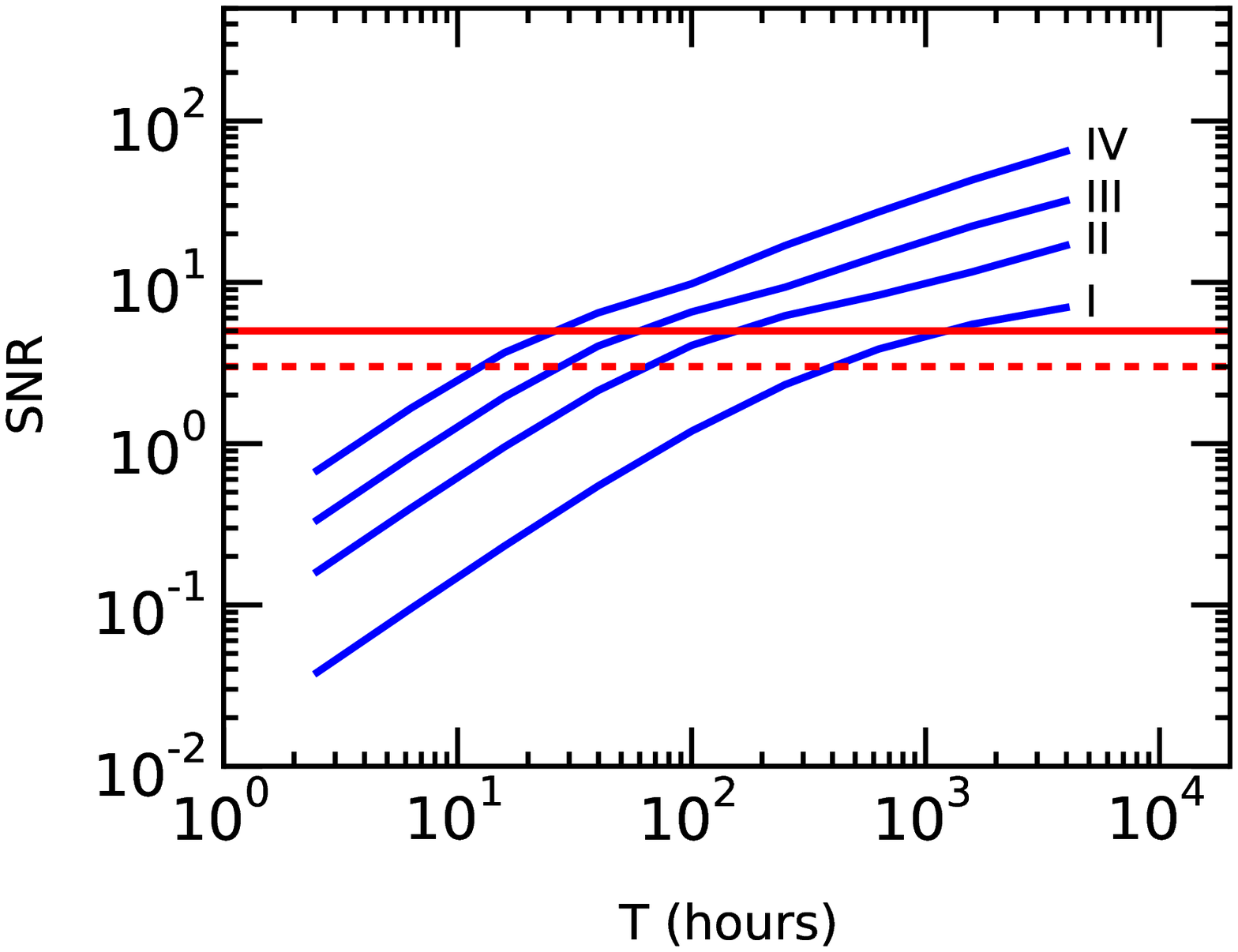}}}
\caption{The Conditional (left)  and Marginalized (right) SNR  for $A_{HI}$
as a function of the observing time  for the  different Phases as indicated in the figure. 
 The horizontal  dashed and solid  lines show  SNR $=3$ and $5$ respectively.}
\label{fig:snr}
\end{center}
\end{figure}
We first focus on estimating  $A_{HI}$  the amplitude   of the HI signal.
The Fisher matrix element $\sqrt{F_{11}}$ gives the   signal to noise ratio (SNR) for  a 
detection of the HI signal ($A_{HI}$) provided the value of $\beta$ is precisely known  
apriori (Conditional SNR). The left panel of Figure  \ref{fig:snr} shows the 
expected Conditional SNR as a function of  the observing time,  and  $t_C$ in 
Table \ref{tab:error} summarizes  the time requirements for  $3-\sigma$ and $5-\sigma$ 
detections.  In reality, the value of $\beta$ is not known apriori, and one hopes to 
measure this from HI observations. While the cosmological parameters which determine
$f(\Omega)$ are known to a relatively high level of accuracy,  there is no direct observational 
handle on the value of $b_{HI}$ at present. It is therefore necessary to allow for the 
possibility that $b_{HI}$ can actually have a value different from $b_{HI}=2$ assumed here. 
A recent compilation of the results from several studies
(Padmanabhan, Roy Choudhury \& Refregier, 2014) has
constrained $b_{HI}$ to be in the range $1.090 \leq b_{HI} \leq 2.06$ 
 in the redshift range $3.25 \leq z  \leq 3.4$. 
In our analysis we  have allowed  $b_{HI}$ to have a value in a 
 larger interval  $1.0 \leq b_{HI} \leq 3.0$, and  we have  marginalized $\beta $ over 
the corresponding interval $0.329 \leq  \beta \leq 0.986$. 
The right  panel of Figure  \ref{fig:snr} shows the 
expected Marginalized  SNR as a function of  the observing time,  and  $t_M$ in 
Table \ref{tab:error} summarizes  the time requirements for  $3-\sigma$ and $5-\sigma$ 
detections.

\begin{table}
\begin{center}
\caption{Here $t_C$ ($t_M$)  is the observing time required for  the Conditional ( Marginalized)
SNR $=3$ and $5$ as respectively indicated in the Table.}
\vspace{.2in}
\label{tab:error}
\begin{tabular}[scale=.3]{|l|c|c|c|c|}
\hline \hline Phase & SNR & $t_C\,({\rm  hr})$ & $ t_M\, ({\rm  hr})$\\ 
 \hline Phase I & 5, 3 & $800,350 $ & $ 1190,390  $ \\ 
\hline Phase II & 5, 3 & $110, 60 $ & $150,70 $ \\ 
\hline Phase III & 5, 3 & $50, 20 $ & $50,20 $\\ 
\hline Phase IV & 5, 3 & $20, 10 $ & $ 25,15 $ \\ 
\hline
\end{tabular}
\end{center}
\end{table}

We find (Figure  \ref{fig:snr})  that for small observing times $(t \le 50 \, {\rm hr})$, 
where the visibilities are dominated by the system noise,  
the Conditional and the Marginalized SNR  both increase as ${\rm SNR} \propto t$.  
The increase in the SNR is slower for larger observing times, and it  
is expected to subsequently  saturate at a limiting value which is set by the cosmic variance 
for very large observing times not shown here. We see (Table \ref{tab:error})
that $\sim 1190 \, {\rm hr}$ of observation 
are needed for a $5-\sigma$ detection with Phase I.  The corresponding  observing time for Phase II 
falls drastically  to $110 \, {\rm hr}$ and $150 \, {\rm hr}$ for  the Conditional and the Marginalized 
cases respectively.  For Phase II, the HI signal is largely dominated by the 
low wave numbers  $k\leq 0.2 \, {\rm Mpc}^{-1}$ (discussed later). Phase I which has a larger antenna spacing and 
smaller frequency bandwidth does not cover many of the low $k$ modes which  dominate the signal contribution for Phase II. 
The required observing times are $\sim 0.5$ and $\sim 0.25$ of those for Phase II for Phases III and IV 
respectively.  The Marginalized SNR are somewhat smaller than the Conditional ones,  the difference 
however is not very large. The required observing time does not differ very much except for Phase II 
where it increases from $110 \, {\rm hr}$ to  $150 \, {\rm hr}$ for a $5-\sigma$ detection.

\begin{figure}[ht]
\begin{center}
\vskip.2cm 
\psfrag{k1}[c][c][0.8][0]{$\Delta A_{HI}/A_{HI}$}
\psfrag{k2}[c][c][0.8][0]{$\Delta \beta/\beta$}
\centerline{{\includegraphics[scale =.33]{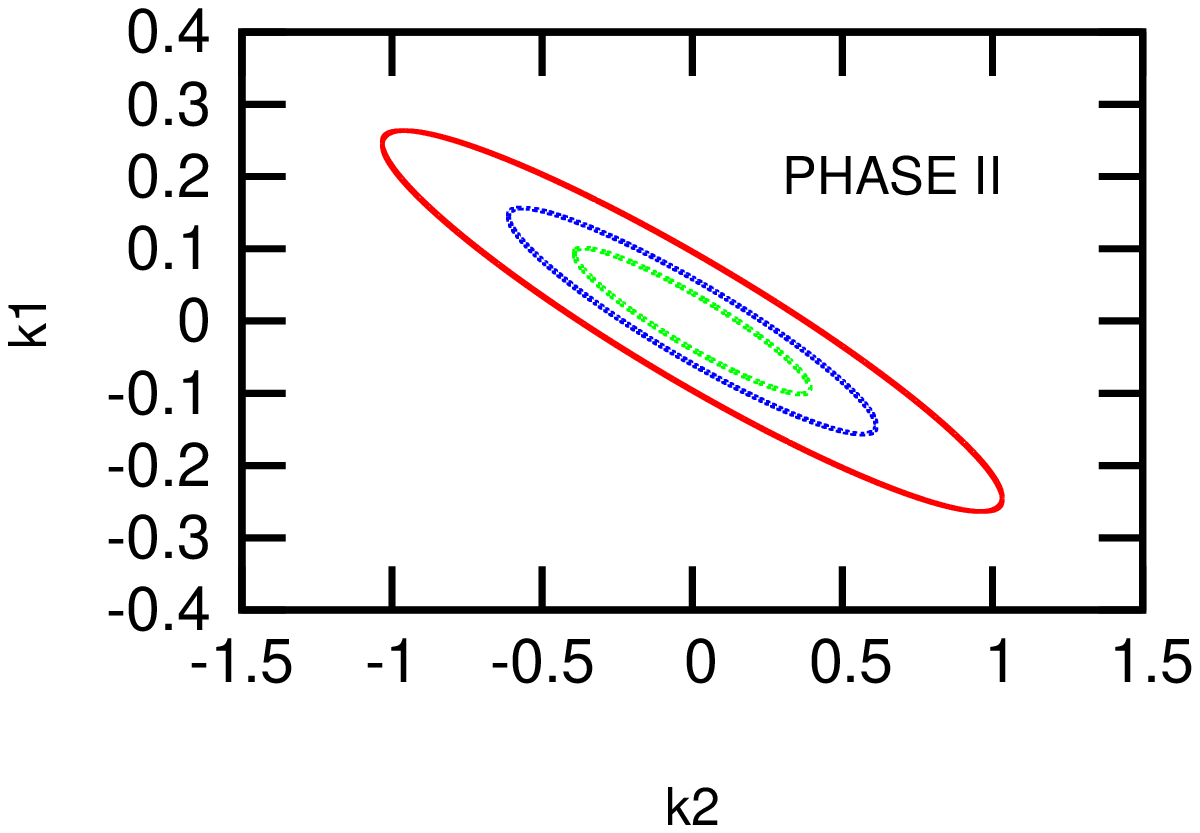}} {
\includegraphics[scale =.33]{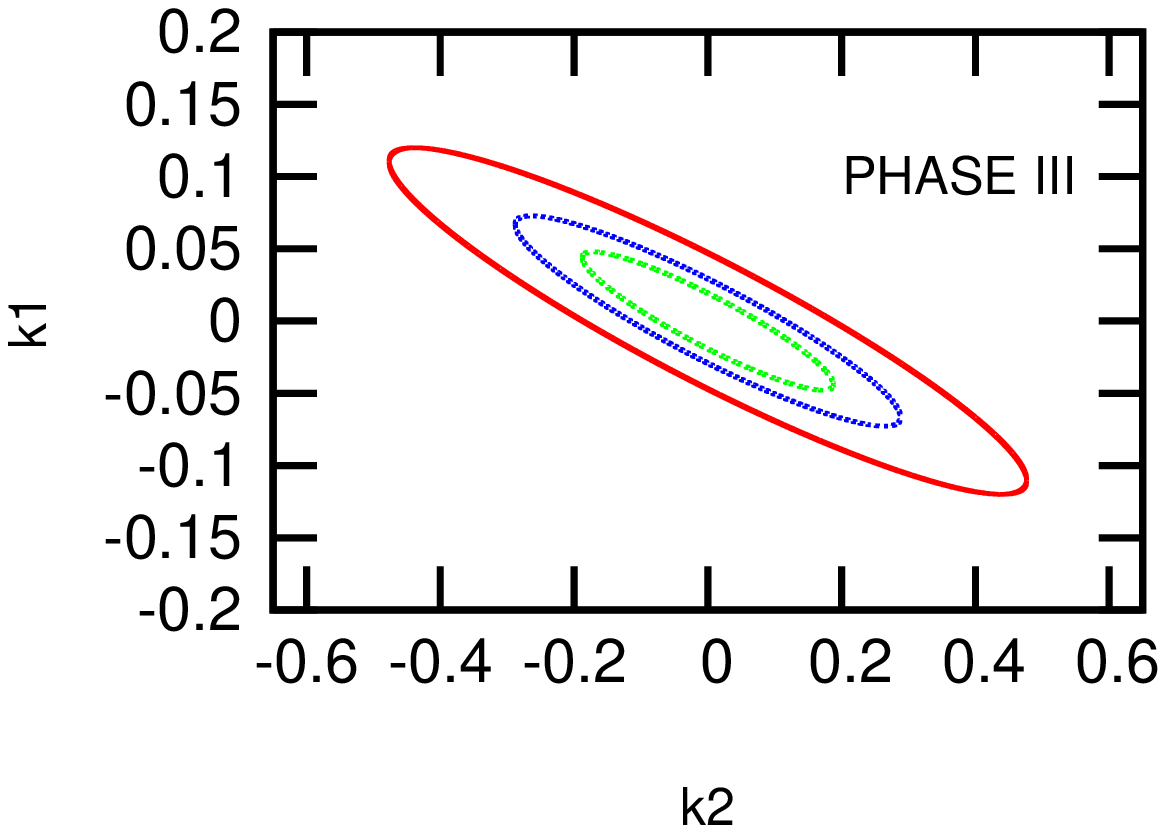}} {
\includegraphics[scale =.33]{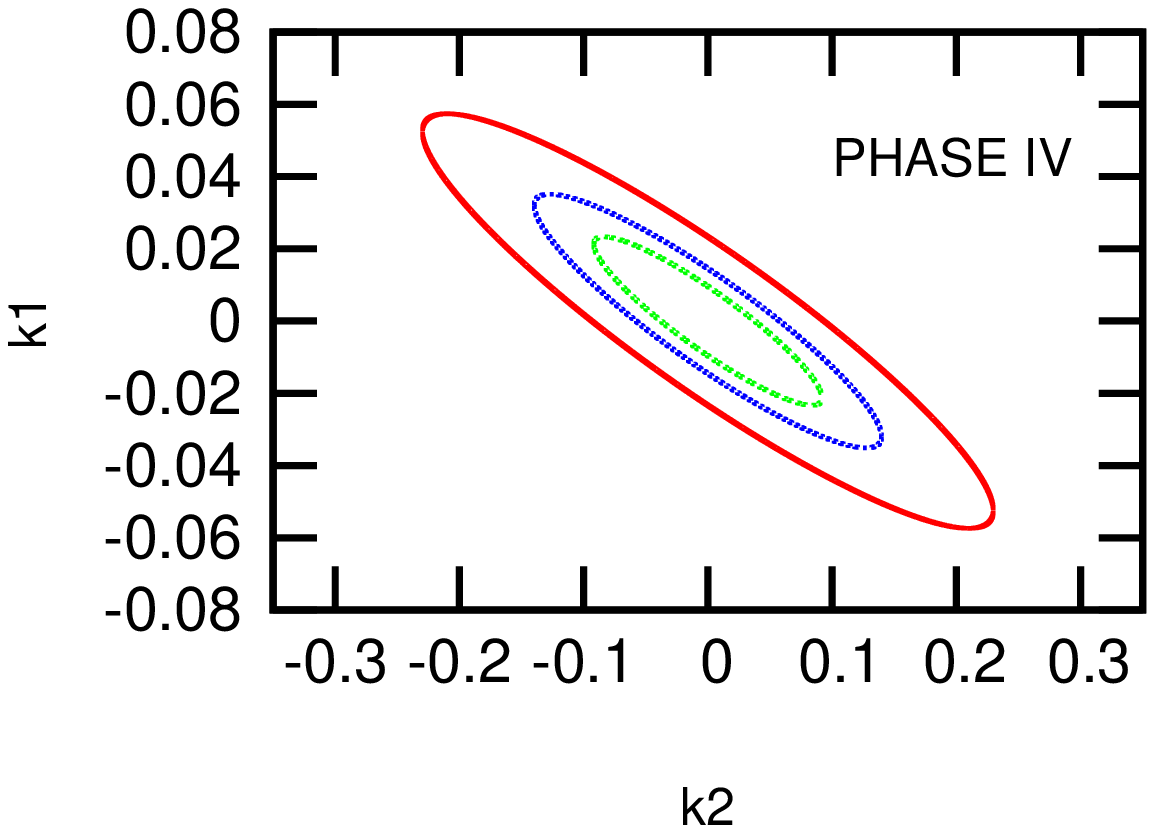}}}
\caption{This shows the expected $1 \sigma$ contours for $\Delta
  \beta/\beta $ and $\Delta A_{HI}/A_{HI} $ with observing time 630
  hrs (outer ellipses), 1600 hours (intermediate ellipses) and 4000
  hours (inner ellipses) respectively for different phases indicated
  in the figures.}
\label{fig:contour}
\end{center}
\end{figure}

We have considered the joint estimation of the two parameters $A_{HI}$ and 
$\beta$ using OWFA. Figure  \ref{fig:contour} shows  the expected $1 \sigma$ confidence
intervals for $\Delta \beta/\beta $ and $\Delta A_{HI}/A_{HI} $ with
three different observing times (630, 1600 and 4000 hr) for 
Phases II, III and IV.  For Phase II, a joint estimation of the parameters
$A_{HI}$ and $\beta$  is possible with 
15\% and 60\% errors respectively using 1600 hr of observation. The errors on the parameters $A_{HI}$
and  $\beta$  for 4000 hr are $\sim 2$ times smaller as compared to 
 1600 hr. 
The constraints are more tight in case of Phases III and  IV.  A 
joint detection of $A_{HI}$ and $\beta$  with 
 3\% and 15\% errors respectively is feasible with 1600 hr of
observation with Phase IV. 

\begin{figure}[h]
\begin{center}
\psfrag{l1}[c][c][1.0][0]{$F_{ab}$}
\psfrag{l2}[c][c][1.0][0]{$k$ Mpc$^{-1}$}
\psfrag{p1}[c][c][0.7][0]{$F_{11}$}
\psfrag{p2}[c][c][0.7][0]{$F_{12}$}
\psfrag{p3}[c][c][0.7][0]{$F_{22}$}
\psfrag{PHASE II}[c][c][0.8][0]{PHASE II}
\psfrag{T=150 hr}[c][c][0.7][0]{$t$=150 hr}
\psfrag{T=800 hr}[c][c][0.7][0]{$t$=800 hr}
\centerline{{\includegraphics[scale =.5]{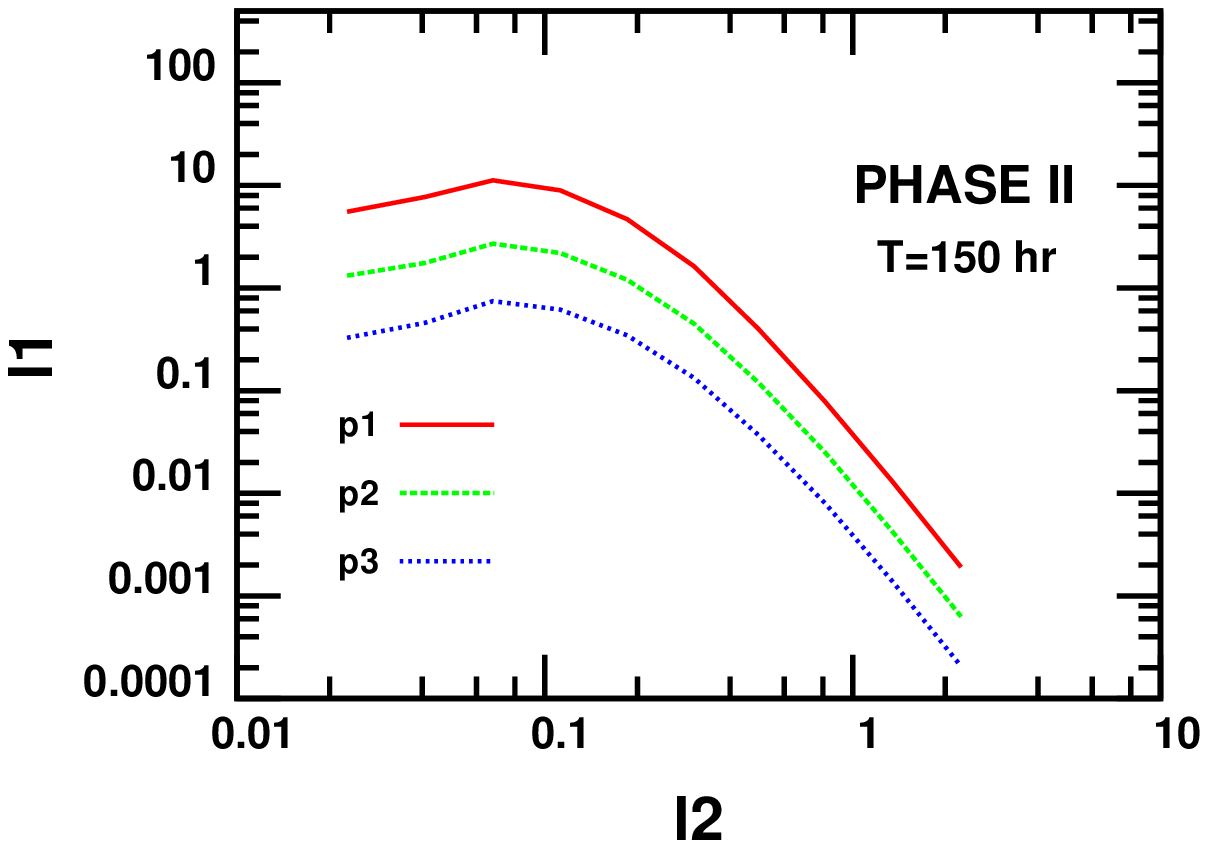}}\hskip0.01cm
{\includegraphics[scale =.5]{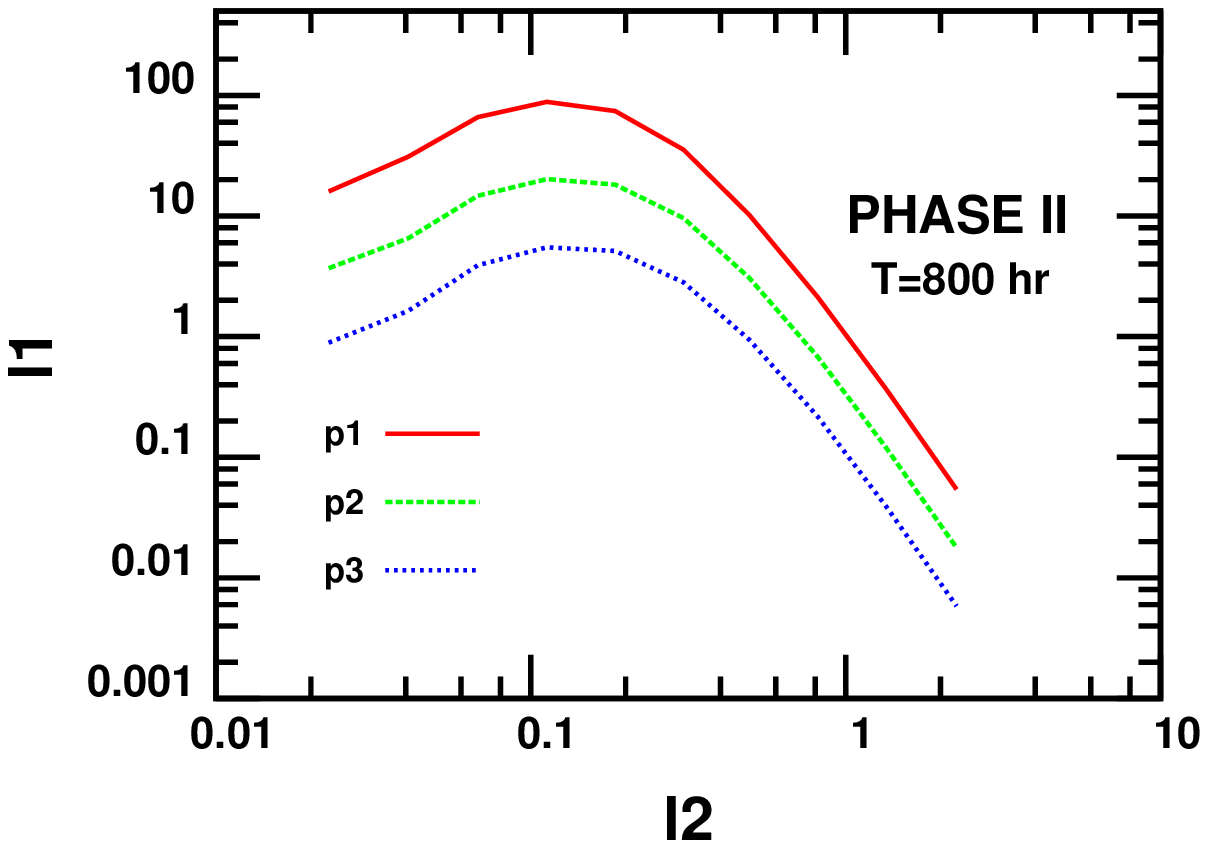}}}
\caption{The relative contribution to the Fisher matrix components $F_{ab}$ 
from the  different k-modes probed by Phase II for 150 hr and 800 hr of observation respectively.}
\label{fig:pkbin}
\end{center}
\end{figure}

It is interesting to investigate  the  $k$  range 
that contribute most to the HI signal at OWFA.  We have seen that 
the Fourier modes $k$ sampled by OWFA are predominantly in the range 
$1 \le k \le 3 \, {\rm Mpc}^{-1}$, and  there are relatively few modes
in the range $ k \le 1  \, {\rm Mpc}^{-1}$ (Figure \ref{fig:pkHI2}). 
However, the HI signal (Figure \ref{fig:convo1}) is much stronger
at the smaller modes, whereas the larger  modes have a weaker HI signal 
 and are dominated by the noise. It is therefore not evident 
as to which $k$ range contributes the most to the OWFA HI signal detection. 
Figure \ref{fig:pkbin} shows the relative contributions to the Fisher matrix 
from the different $k$ modes. We see that for $t=150 \, {\rm hr}$, which 
corresponds to a $5-\sigma$ detection, the bulk of the contribution is
from the range $ k \le 0.1  \, {\rm Mpc}^{-1}$. The larger modes do not 
contribute much to the signal.  We have also considered $t=800 \, {\rm hr}$.  
Here we have a slightly larger range $k \le 0.2  \, {\rm Mpc}^{-1}$ and 
the contribution  peaks around  $k \approx  0.1  \, {\rm Mpc}^{-1}$. 
In a nutshell, the OWFA HI signal is predominantly from the $k$ range 
$0.018  \le k \le 0.2  \, {\rm Mpc}^{-1}$. The larger modes, though 
abundant, do not contribute much to the HI signal.

\section{Summary and conclusions}
\label{sec:sum}
We have considered  four different Phases of OFWA, and 
studied the prospects of detecting the  redshifted 21-cm  HI  signal at  $326.5 \, {\rm MHz}$ 
which corresponds to redshift $z = 3.35$.
Phases I and II are currently under development and are expected to be functional in the near future. 
Phases  III and IV are two hypothetical  configurations which have been considered as 
possible future expansions. We have used the Fisher matrix analysis to predict  the accuracy with 
which it will be possible to estimate the two parameters  $A_{HI}$ and $\beta$ using OWFA. 
Here $A_{HI}$ is the amplitude of the 21-cm brightness temperature power spectrum and $\beta$ is the 
linear redshift space distortion parameter.  For the purpose of this work we make the most optimistic assumption that the  foreground contributions are not changing  across  different frequency channels, and hence they only contribute to the $\kp=0$ mode. In reality the foreground contamination will extend to other modes also. Further, the chromatic response of the interferometer, calibration errors, systematics in the receivers and radio-frequency interference (RFI) have not been considered in the paper.

Focusing first  on just detecting the HI signal, we have marginalized $\beta$ and considered the 
error estimates on $A_{HI}$ alone. We find that a $5-\sigma$ detection of the HI signal is possible
with $1190$ and $150 \, {\rm hr}$ of observation for Phases I and II respectively.  The observing 
time is reduced by factor $\sim 0.5$ and $\sim 0.25$ compared to Phase II for Phases III and IV respectively. 
We find that there is a significant improvement in the  prospects of a detection using Phase II as
compared to Phase I, and we have mainly considered  Phase II  for much of the discussion in the paper. 

We have also considered the  joint estimation of the parameters $A_{HI}$ and $\beta$. 
 For Phase II, a joint estimation of the parameters
$A_{HI}$ and $\beta$  is possible with 15\% and 60\% errors respectively using 1600 hr of observation. 
To estimate $\beta$ it is necessary to sample Fourier modes $\k$ of  a fixed magnitude $k$ 
which are oriented  at different directions to the line  of sight. In other words, $\mu=\kpar/k$ 
should uniformly span the  entire range $-1 \le \mu \le 1$. However, the $\kpar$ values are
 much larger than $\kperp$, and the Fourier modes are largely concentrated around $\mu \sim 1$
, for Phase II  (Section \ref{sec:vc}).  The restriction  arises from the limited 
OWFA frequency bandwidth (Table~\ref{tab:array}) which is restricted by the anti-aliasing filter.

Multi-field observations and larger bandwidth ($> 30 \,{\rm MHz})$ of the OWFA  hold the potential to  probe of the expansion history and  constrain cosmological parameters  using BAO measurements from the large-scale HI fluctuations at $z = 3.35$. Anisotropies in the clustering pattern in redshifted 21-cm maps at this redshift  produced by Alcock-Paczyski effect has the possibility of probing cosmology and structure formation. It is also possible to  constrain neutrino masses of using OWFA and compare among different
fields of cosmology (LSS, CMBR, BBN). Thus the OWFA could provide highly complementary constraints on neutrino masses. We leave investigation of such issues for future studies.

The present work has assumed that the shape of the HI power spectrum 
is exactly determined by the $\Lambda$CDM model, and has only  focused
estimating the overall amplitude $A_{HI}$ from OWFA observations. 
The OWFA HI signal is predominantly from the $k$ range 
$0.02  \le k \le 0.2  \, {\rm Mpc}^{-1}$. It is possible 
to use OWFA observations to  estimate $P_{HI}(k)$ 
in several separate  bins over this $k$ range, without assuming the 
anything about the shape of the HI power spectrum. In a forthcoming paper, 
we plan to calculate  Fisher matrix estimates for the binned power 
spectrum.  

\section*{Acknowledgment}
The authors acknowledge Jayaram N. Chengalur, Jasjeet S. Bagla, Tirthankar
Roy Choudhury, C.R. Subrahmanya, P.K. Manoharan and Visweshwar Ram
Marthi for useful discussions. AKS would like to acknowledge Rajesh Mon-
dal and Suman Chatterjee for their help. SSA would like to acknowledge
CTS, IIT Kharagpur for the use of its facilities and the support by DST,
India, under Project No. SR/FTP/PS-088/2010. SSA would also like to
thank the authorities of IUCAA, Pune, India for providing the Visiting As-
sociateship programme.

\end{document}